\documentstyle[12pt,sprocl]{article}

\begin{document}

\title{
$R$-Parity Violation and the HERA Events.
      }

\author{
Sreerup Raychaudhuri
        }

\address{
Theory Division, CERN, CH 1211 Geneva 23, Switzerland.
        }

\maketitle

\abstracts{
The excess in high-$Q^2$ events at HERA is introduced and 
possible explanations within the framework of $R$-parity
violating supersymmetry are discussed.
          }

Early this year, the H1 and ZEUS Collaborations at the 
HERA facility at DESY, Hamburg, created a sensation by 
announcing~\cite{H1ZEUS} that they had an excess in 
back-scattered positrons (in $e^+ p$ collisions) over the 
deep inelastic scattering (DIS) predictions of the Standard 
Model (SM). Though statistics were low, it appeared to be a 
hint of new physics beyond the SM, and a series of theoretical 
papers investigating the effect followed. Subsequently, the 
two collaborations presented~\cite{LP97} the results of later 
runs, in which they again had an excess, though at a lower 
level. More data have been collected in the subsequent run of 
HERA, which are still being analysed. This is the current 
situation. Thus, the case for new physics at HERA --- though 
weaker than it was initially thought to be --- still remains 
the best experimental hint of physics beyond the SM.

The scattering of positrons (beam energy $E^0_e = 27.5$ GeV) 
from protons (beam energy $E^0_p = 820$ GeV) at HERA can lead 
to neutral current (NC) final states with a positron ($e^+$) 
and a jet ($J$) describable by any two of the following
observables: $E_e, \theta_e, E_J, \theta_J$ where $E_e,E_J$ are 
the energies and $\theta_e, \theta_J$ are the scattering angles 
respectively (with respect to the proton beam direction). 
The observed excess can be clearly seen in Fig.1($a$), which shows
the measured cross-section as a function of a cut on the minimum 
$Q^2$ of the events. As errors are large, the excess is at the 
$2\sigma$ level, which means that they could well be due to 
statistical fluctuations. It is also clear that the H1
Collaboration sees a larger deviation from the SM than the ZEUS 
Collaboration.
 
\begin{figure}[htb]
\vskip 3.0in
      \relax\noindent\hskip -0.6in\relax{\includegraphics{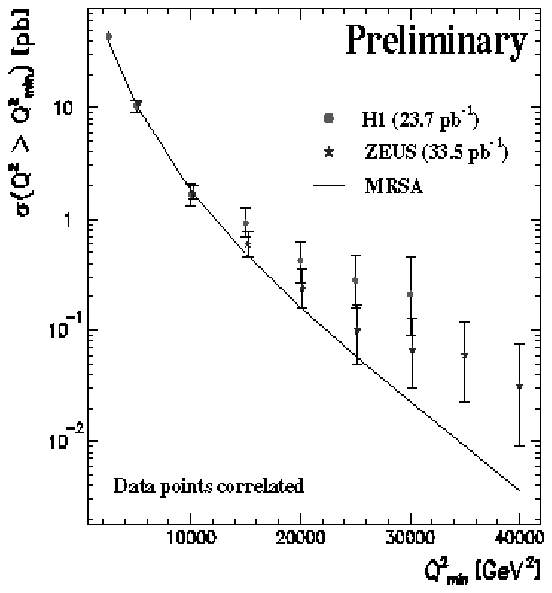}}
      \relax\noindent\hskip  2.7in\relax{\includegraphics{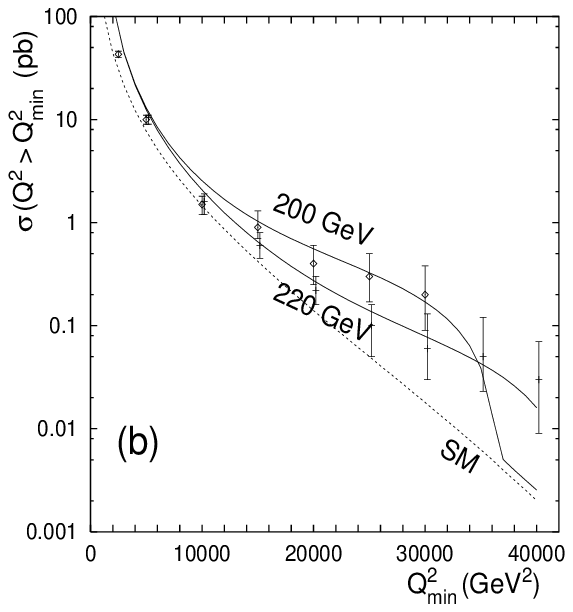}}
\vskip -0.8in
\caption{\footnotesize\sl 
($a$) Comparison of SM Prediction with data (from Ref.2). The solid 
line shows the DIS prediction. ($b$) Effects of a resonance of mass 
200 GeV and 220 GeV (solid lines) for $\lambda'_{1i1} 
\sqrt{B.R.(e^+J)} = 0.04$.}
\end{figure}

It is, of course, possible to dismiss the excess as an artefact.
Nevertheless, it is interesting to seek theoretical explanations 
of the effect, in case it should turn out to be a genuine one. 
Three kinds of solution have, in fact, been tried. These are 
($a$) attempts to modify the proton structure functions~\cite{sfun}, 
($b$) explanations in terms of contact interactions arising out of 
new physics at a high scale~\cite{contact}, and ($c$) the postulate 
that a new particle resonance, of mass around 200 GeV, is causing 
the effect. The first two options have been shown not to work 
well~\cite{sfun}, and it is, thus, to the third possibility that we 
must appeal. A resonance produced at HERA in a $e^+ q$ or $e^+ g$ 
collision must have the quantum numbers of a leptoquark or a 
leptogluon respectively. As solutions with vector leptoquarks or 
leptogluons do not fit the kinematic profiles of the observed events 
well, a better option is to postulate a {\em scalar leptoquark} 
resonance. Such particles are predicted in various extensions of the 
SM, of which one is the $R$-parity-violating version of the minimal 
supersymmetric extension of the SM, where the squarks can behave as 
leptoquarks.

The Lagrangian describing this last interaction is given by:
\begin{equation}
{\cal L} = \lambda'_{ijk} 
\left[ \bar{\ell}_i P_R d_k \widetilde{u}^*_{Lj} 
~-~ (i \leftrightarrow j) \right] ~+~ {\rm H.c.}
\end{equation}
where $P_R = \frac{1}{2} (1 + \gamma_5)$ and $\widetilde{u}_{L}$ 
is the scalar superpartner of the left-chiral $u$-quark. The 
$i,j,k$ are generation indices. If we wish to obtain reasonably 
large cross-sections, the only relevant couplings are 
$\lambda'_{121}$, $\lambda'_{131}$, $\lambda_{132}$ corresponding 
to $e^+ d \rightarrow \widetilde{c}_L$, 
$e^+ d \rightarrow \widetilde{t}_L$, and
$e^+ s \rightarrow \widetilde{t}_L$ respectively~\cite{RPVsolution}. 
(For the last option, it is a sea $s$-quark which contributes: the 
flux being small, one requires a large $\lambda'_{132}$ to get the 
observed excess~\cite{ElLoSri}.) The calculation is simple and 
Fig.1($b$) shows the results as a function of the minimum $Q^2$. 
It is obvious that the data are compatible with resonant masses in 
the 200--220 GeV range, with H1 data preferring the lower and ZEUS 
data preferring the upper mass value. There will also be some 
variation with the coupling strength, which is not shown in the 
figure. One must hope that more data will lead to a better idea of 
the mass.

An intensive search has also been carried out at the Tevatron for 
a pair of leptoquarks (or squarks), which can be produced through 
the same interaction as the above. The signal would be a dielectron 
and one or two jets.  These searches have proved negative~\cite{CDFD0}, 
leading to the bound $m_{\widetilde q} > 240$ Gev for branching ratio 
$B.R.(e^+ J ) = 1$, 206 GeV for $B.R.(e^+ J) = 0.5$. Thus, if the 
HERA events are to be explained by a leptoquark/squark resonance of 
mass around 200 GeV, this particle must have a $B.R.(e^+ J)$ of 0.4 
or less. This is possible for a charge-$\frac{2}{3}$ leptoquark if 
it mixes with another leptoquark, but such models look a little 
contrived. On the other hand, a squark can
easily have $R$-parity {\em conserving} decay modes to gaugino
states. In a sense, therefore, the negative results of the leptoquark
searches at the Tevatron make supersymmetry the most attractive
explanation of the HERA 
excess. Another interesting possibility~\cite{DebSree} -- that of 
seeing like-sign dileptons from gluino pair-production in the event 
of a ${\widetilde c}_L$ resonance around 200 GeV -- has been 
investigated by the CDF Collaboration with negative 
results~\cite{Kamon}; this leads to fresh bounds on the gluino mass 
if the HERA events have a genuine supersymmetric solution.

The question which immediately arises as a result of the above 
reflections is: are the other decay channels observable ? 
The most likely other decay channel, to $\bar \nu_e + J$, 
must lead, at HERA, to a corresponding excess in the charged current 
(CC) DIS data. In fact, the H1 Collaboration {\it has} found
a tentative CC excess; the ZEUS Collaboration has practically 
none~\cite{LP97}. As the $\bar \nu_e J$ decay
mode will have a branching ratio of 0.6 or more, one must  
postulate low detection efficiencies, in which case
a mixed leptoquark solution runs into difficulties. In supersymmetry, 
however, it is possible to have scenarios with low efficiencies.
One of these~\cite{AlGiMan} assumes a
light sneutrino, of mass 60 -- 80 GeV, in addition to a 
${\widetilde c}_L$ squark (with a $\lambda'_{121}$
coupling). Unfortunately, this scenario also predicts an
excess of four-jet events at LEP, which has not been
found (despite early excitement over the ALEPH Collaboration's
results). Another suggestion~\cite{KonKob} that a resonant 
${\widetilde t}_L$ may directly decay into a 
${\widetilde b}_L$ (which goes to $\bar \nu_e d$ through 
$\lambda'_{131}$) leads to
problems with the number of jets and is rather difficult to fit in 
with $\rho$-parameter constraints.

Two possible solutions~\cite{4authors} which do not have these drawbacks
require a ${\widetilde t}_1$ resonance, where 
$m_{\widetilde t_1} - m_{\widetilde b_L} < 35$ GeV and
$m_{\widetilde t_1} > m_{\widetilde \chi^+_1} > m_{\widetilde b_L}$,
and a mixed stop state is required to have the small splitting
in stop-sbottom masses. It is worth mentioning that there are 
theoretical drawbacks associated with a light ${\widetilde c}_L$
which do not apply~\cite{Joshipura} in the case of a ${\widetilde t}$. 
One could now have ($a$) a neutralino $\chi_1^0$
lying between the $\chi_1^+$ and the ${\widetilde b}_L$ in mass,
which leads to the decay chain 
${\widetilde t}_1 \rightarrow b {\widetilde \chi}^+_1
\rightarrow b f \bar f' {\widetilde \chi}^0_1
\rightarrow b f \bar f' b {\widetilde b}_L
\rightarrow b f \bar f' b \bar \nu d$. 
For the assumed mass-spectrum most of the jets resulting from the
cascade decays are soft and would pass undetected. Single hard jets
would result from the last decay and the observed level of 
excess CC events can be explained. Alternatively ($b$) the chargino 
produced in the first decay can undergo the
Cabibbo-Kobayashi-Maskawa-suppressed decay 
${\widetilde \chi}_1^+ \rightarrow c {\widetilde b}_L$ 
where ${\widetilde b}_L \rightarrow \bar \nu d$ as before.
Once again, one can have single jet events, though 
one requires rather large values of the 
soft supersymmetry-breaking parameter $\mu$ 
to obtain the observed CC excess. Fortunately, 
such values do not contradict current low-energy bounds on 
supersymmetric models.
 
To conclude, a great deal of excitement has been generated by the
HERA excess for which it seems that $R$-parity
violating supersymmetry is the best explanation.
With the present level of statistics it is hard to make a more definite 
statement. Further results from HERA are, therefore, eagerly awaited.

\end{document}